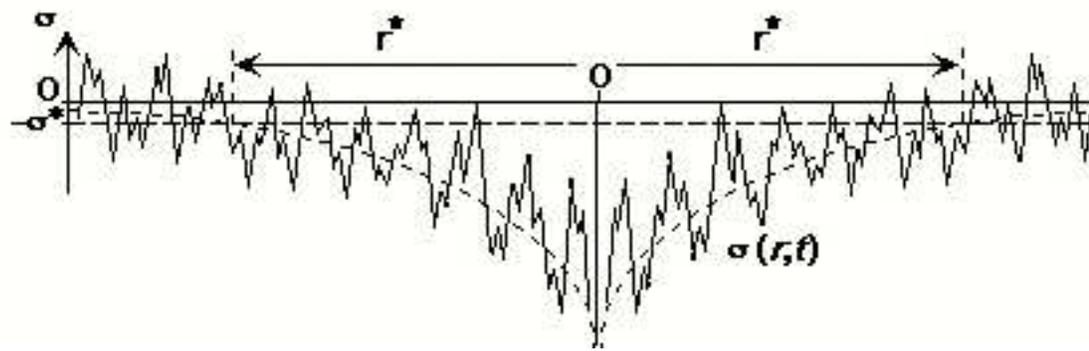

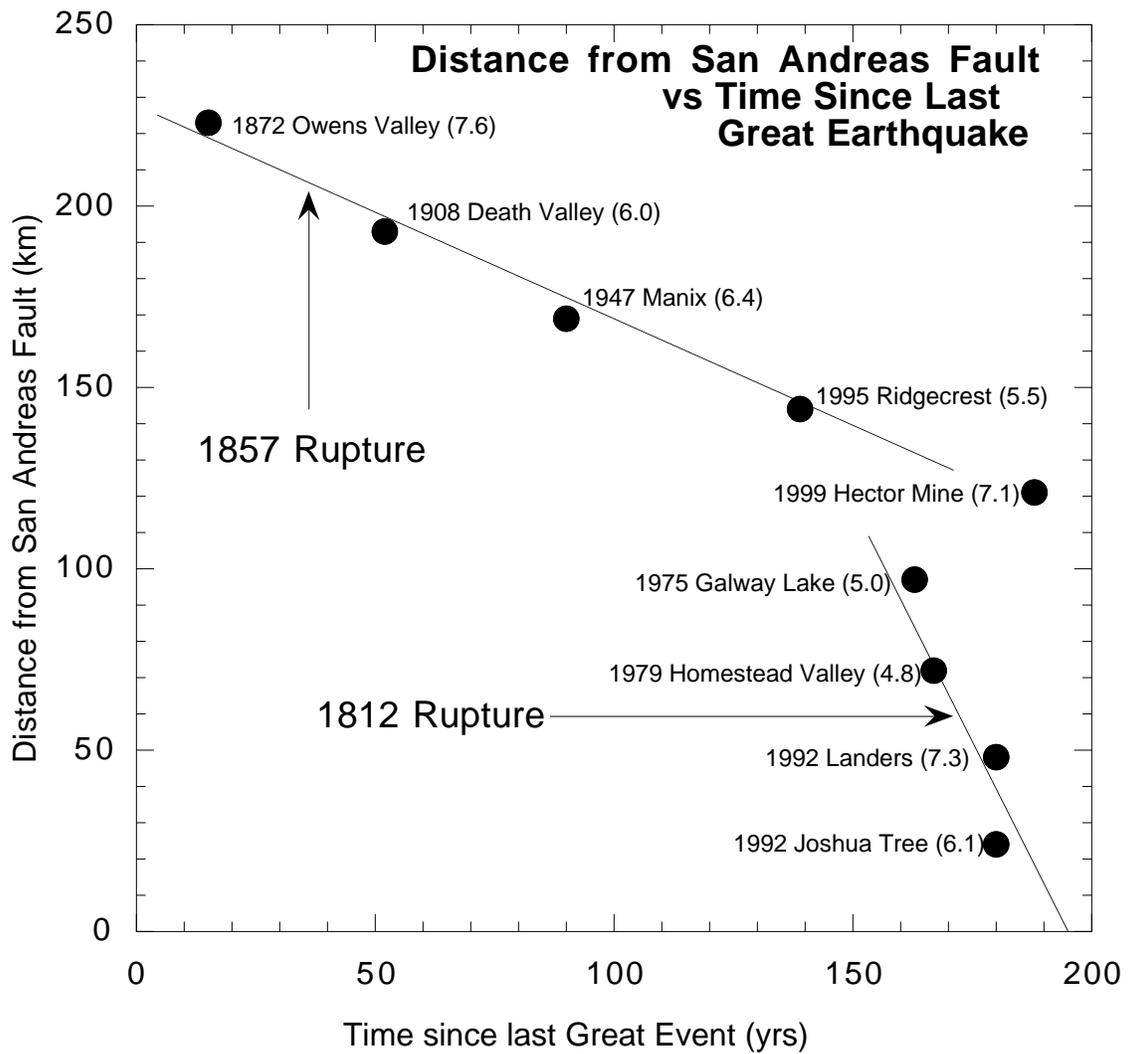

# Positive Feedback, Memory and the Predictability of Earthquakes

C.G. Sammis[1] and D. Sornette[2,3]

[1]Department of Earth Sciences
University of Southern California
Los Angeles, CA 90089-0740

[2]Institute of Geophysics and Planetary Physics
and Department of Earth and Space Science
University of California, Los Angeles, California 90095

[3]Laboratoire de Physique de la Matière Condensée
CNRS UMR6622 and Université des Sciences, B.P. 70, Parc Valrose
06108 Nice Cedex 2, France

**Abstract**: We review the "critical point" concept for large earthquakes and enlarge it in the framework of so-called "finite-time singularities". The singular behavior associated with accelerated seismic release is shown to result from a positive feedback of the seismic activity on its release rate. The most important mechanisms for such positive feedback are presented. We introduce and solve analytically a novel simple model of geometrical positive feedback in which the stress shadow cast by the last large earthquake is progressively fragmented by the increasing tectonic stress.

**Introduction**

Are earthquakes predictable? The answer, of course, depends on what's meant by a prediction. In the broadest sense, the plate tectonics paradigm makes predictions. It predicts that earthquakes are far more likely to occur at the boundaries between plates than within their interiors. Actually, plate tectonics theory was in part based on this "in-sample" observation, which is verified continuously "out-of-sample". It also predicts an overall rate to the process. Averaged over time, the summed moments of the earthquakes is consistent with the relative motion between the plates determined from the analysis of magnetic anomalies, correcting for aseismic visco-plastic deformations.

The forecasting of individual large events has been more problematical. While the paleoseismological dating of large prehistoric earthquakes has confirmed the plate tectonics hypothesis, the timing between individual events is extremely erratic. For example, trenching by Sieh et al. (1989) found that the average recurrence interval for the last 10 large earthquakes on the San Andreas Fault north of Los Angeles is about 132 years. Since the long-term slip rate on the Southern San Andreas is about 3 cm/yr, this corresponds to an average displacement of about 4 m per large earthquake – a very reasonable value. The problem is that the intervals between events range from 44 to 332 years. This lack of quasiperiodicity in large events is also evident in the failed Parkfield prediction where the current interval is approaching 35 years despite a prior string of intervals averaging about 21 years (see, e.g., Bakun and McEvilly, 1984). These observations have dimmed the hope that large earthquakes can be forecast based solely on the past history of large events on the same fault.

An alternative forecasting strategy is based on physical precursors observed to occur just before macroscopic failure in the laboratory. Most of these are associated with microfracture damage and the associated dilatancy observed to precede the formation of a macroscopic shear failure of rock specimens under compressive loading. These laboratory observations have been incorporated into the "dilatancy-diffusion model" for earthquakes (Nur, 1972; Whitcomb et al., 1973; Scholz et al., 1973; Griggs et al., 1975). However, the search for physical precursors before large earthquakes has been disappointing. The high hopes raised by the reports of Chinese success in using physical precursors to forecast the 1975 Haicheng earthquake have dissipated with the worldwide failure to produce additional valid predictions (for a review, see Turcotte, 1991).

In the USA, current work on earthquake prediction is primarily based on the search for precursors to large events in the seismicity itself. One motivation comes from a statistical physics interpretation of regional seismicity as being characteristic of a system at or near a statistically stationary dynamical critical point dubbed self-organized criticality (see Sornette (2000) for a review of mechanisms). Such a self-organized critical state is characterized by power law distributions of event sizes and long-range spatial correlation (Majumdar and Dhar, 1991) of fluctuations around the statistically stationary state. Since earthquakes are characterized by several power laws, such as the Gutenberg-Richter frequency-magnitude relation, Omori's Law for aftershocks and the fractal distribution of hypocenters, the application of this concept of self-organized criticality to earthquakes is now often taken for granted in the seismological community.

However, the implication of this self-organized critical (statistically stationary) state for the predictability of large earthquakes remains controversial. Geller et al. (1997) pointed out that if the crust follows the strict tenets of self-organized criticality, then earthquakes are inherently unpredictable. According to this view point, simple self-organized critical systems having short range interaction and conserved stress (except at the boundaries) tend to sit at or near the statistically stationary dynamical critical point such that the probability for a small event to cascade into a large one is not preconditioned by previous events. Physically speaking, this is because the long-range stress correlation is not significantly destroyed by large events in such systems. This view has actually been proved wrong by direct tests of the predictability of large events in several sandpile models which show detectable correlations between precursory smaller events and the largest events (Pepke and Carlson, 1994), though the correlations may be very weak in some cases. The physical reason lies in the fact that a statistically stationary dynamical critical state is in fact the result of subtle and long-range correlations in space and time between events. As a consequence, a large event occurs on a long-range correlated landscape. In practice however, these effects are small and hard to detect and quantify precisely.

The alternative point of view is that large system-wide events produce significant perturbations that move a system away from the statistically stationary dynamical critical point. Non-conservative automata (see, e.g., Olami et al., 1992) or automata having correlated heterogeneity on the scale of the network have been shown to produce such perturbations (Huang et al., 1998; Sammis and Smith, 1999). Because these systems have a strong memory of past events, prediction becomes much more possible. Pushing this concept, Sornette and Sammis (1995) have proposed that great earthquakes are themselves "critical points", i.e., the culmination of a non-stationary precursory activity accelerating up to the critical point. We emphasize that this concept is fundamentally different from the long-time view of the crust as evolving spontaneously in a statistically stationary critical state, called self-organized criticality (SOC). In the SOC view, all events belong to the same global population and participate in shaping the self-organized critical state. By contrast, in the critical point view a great earthquake plays a very special role and signals the end of a cycle on its fault network. The dynamical organization is not statistically stationary but evolves as the great earthquake becomes more probable. As we said, predictability becomes possible and may be implemented by monitoring the approach of the fault network toward a critical state.

The basic tenets of the critical point model for regional seismicity can be simply stated as follows:
- A large (network sized) earthquake is only possible when the crust has reached a critical state. This is because highly stressed patches must be correlated at the scale of the fault network if an event, once nucleated, is to grow by jumping geometrical and rheological barriers.
- A large earthquake moves portions of the network out of the critical state by destroying stress correlation. This produces the observed shadow zones for intermediate size events.
- Tectonic loading combined with the stress transfer from smaller events reestablish long-range stress correlation thus making the next large event possible.

Note that nothing in this hypothesis implies that the seismic cycle, thus defined, has to be periodic. Presumably, observed variations in recurrence interval discussed above result from details of the stress transfer from smaller events and from details of the nucleation process. Also, nothing in the hypothesis says that the large earthquake has to occur when the system reaches the critical state – only that a large earthquake is then possible. The exact time of the large earthquake may depend on details of the nucleation process, which may itself have significant time dependence (see, e.g., Dieterich, 1992). Predictability in this model comes from monitoring the approach of the fault network back to the critical state.

Observations of the acceleration of seismic moment leading up to large events and "stress shadows" following them have been interpreted as evidence that seismic cycles represent the approach and retreat of a fault network for a critical state. Harris and Simpson (1996) and Jones and Hauksson (1997) present observational evidence for "Coulomb-stress shadows" following the 1857 Fort Tejon, the 1952 Kern County, and the 1992 Landers earthquakes. Harris and Simpson found no M>5.5 earthquakes in the Coulomb-stress shadows for 50 years following the 1857 and 1952 events, while M>5.5 events continued at a normal rate in areas where the Coulomb stress was not lowered. This lack of intermediate sized events is the expected result of a reduction in stress correlation length associated with a retreat from the critical state following a large event.

The fundamental hypothesis of the critical point model is that fault networks naturally evolve back toward the critical state following a large event, and that this approach to criticality can be monitored using regional seismicity. Three approaches to this monitoring have been proposed. The first is to monitor the increase in event size as a proxy for the growing correlation length. Sykes and Jaumé (1990) documented this effect as an increase in intermediate-sized events in a large region surrounding the impending large earthquake. Bufe and Varnes (1993) quantified the effect of these intermediate-sized events in terms of a power law time-to-failure equation of the form

$$\sum_{j=1}^{N(t)} E_j^{1/2} = A + B(t_c - t)^m \qquad (1)$$

where $E_j$ is the energy of the $j^{th}$ event, $t_c$ is the time that the system reaches the critical state and a large event becomes possible. The quantity $\sum_{j=1}^{N(t)} E_j^{1/2}$ is usually referred to as the "cumulative Benioff strain". The parameter $B$ is negative and $m<1$ (typically $m \approx 0.3$) so that the cumulative Benioff strain accelerates up to the finite value A with a diverging slope as $t$ goes to $t_c$. In practice, only the intermediate sized events (typically two magnitude units below the large event or greater) are included in the sum (1) since the Benioff strain is dominated by small events.

Sornette and Sammis (1995) pointed out that equation (1) is the expected behavior near a critical point and proposed in addition to enrich it with log-periodic corrections (see also Saleur et al, 1996). Bowman et al. (1998) optimized the fit to equation (1) with respect to the size of the critical region to find a simple scaling relation between magnitude and radius for circular regions. In fact, they found the same scaling proposed by Keilis-Borok and Malinovskaya (1964) based simply on the width of the fault network. A good review of this line of research is in Jaumé and Sykes (1999).

More recently, a second method has been proposed by Zoller et al. (2001) who estimate the growing correlation length directly from the earthquake catalog using single-link cluster analysis. Their method has the advantage of working with smaller events thus improving the statistical significance of the results.

A third method for monitoring the approach to criticality is based on the notion that a system approaching the critical state becomes increasingly sensitive to tidal stress perturbations. Yin et al. (1995) proposed that a correlation between solid earth tides and seismicity develops locally just before a large earthquake. They quantify this effect by calculating the average seismic energy release in a region surrounding the impending earthquake during periods when the effective tidal stress on the fault plane is positive divided by the energy released during periods when it is negative. Dubbed the Load/Unload Response Ratio (LURR), Yin et al. (1995, 2001) reported LURR calculations for a number of earthquakes in China, Australia, and California, and suggest that this ratio increases prior to the large earthquakes that they studied. The time period over which this increase occurs appears to be several months to a year or two. Recently, Yin (2001) has broadened their rationale to encompass the critical point model for large earthquakes. They support this interpretation with examples which illustrate that the largest and best defined anomalies in LURR seem to occur when the choice of the region over which seismicity is averaged follows the scaling law developed by Bowman et al. (1998) using the power law acceleration of regional seismicity. In a related development, Wang et al. (2000) carried out experiments with the Lattice Solid Model of Mora and Place (1994, 1998) in which sinusoidal perturbations to the loading force were applied in their numerical simulations. The results showed a confirmation that anomalous values of LURR occurred prior to large events in the model.

There are, however, theoretical arguments that the LURR effect should not be observable. Dieterich, (1987, 1992) points out that fault patches with rate- and state-dependent friction properties are characterized by an interval of self-driven accelerating slip prior to instability. This delayed nucleation provides an explanation for the time delay between the stress change attending an earthquake and the resultant aftershocks, and even explains why the rate of aftershocks decays as 1/time. Dieterich (1987) investigated the behavior of a population of patches having a rate- and state-dependent friction rheology when they are subjected to loading with a periodic component. He found that the response to tidal loading should not be detectable in a homogeneous crust at normal stresses exceeding 8 MPa (or equivalently, at depths greater than about 300 m). It is not clear that any enhancement in correlation associated with the approach to criticality will not likewise be obscured by delayed nucleation. It should be noted that the Lattice Solid Model does not yet include rate- and state-dependent friction and the associated delayed nucleation and thus the results of Wang et al. (2000) can not be taken as theoretical verification of the LURR effect.

Ouillon and Sornette (2000) have performed tests of the concept that large earthquakes are "critical points" on rockbursts in deep South African mines. First, using the simplest signature of criticality in terms of a power-law time-to-failure formula, they find evidence both for accelerated seismicity and for the presence of log-periodic behavior (initially proposed by Sornette and Sammis (1995)) in the cumulative Benioff strain with a preferred scaling factor close to 2. They have also proposed a new algorithm based on a space and time smoothing procedure, which is also intended to account for the finite range and finite duration of mechanical interactions between events. This new algorithm provides a much more robust and efficient construction of the optimal correlation region, which allows them the use of the log-periodic formula directly in the search process. A preliminary test
on the largest event on the catalog shows a remarkable good quality with a dramatic improvement in accuracy and robustness. This suggests new potential for improving on the simple power law fits by using log-periodic signals.

Until recently, this critical point hypothesis has been a conceptual model based on the analogy with phase transitions. Theoretical support has come from simple models such as cellular automata, with (Anghel et al., 1999; 2000; Sá Martins et al., 2001) and without (Huang et al., 1998; Sammis and Smith, 1999) long-range interaction and granular simulators (Mora et al., 2000; Mora and Place, 2001). Models of regional seismicity with more faithful fault geometry have been developed that also show accelerating seismicity before large model events (Heimpel, 1997; Ben-Zion and Lyakhovsky, 2001; and King and Bowman, 2001).

In their cellular automaton model of earthquake faults, Anghel et al. (2000) find that the scaling of earthquake events in models of faults with long-range stress transfer is composed of at least three distinct regions, corresponding to three classes of earthquakes with different underlying physical mechanisms. The largest "breakout" events are found to be associated with a spinodal critical point. In this model, Sá Martins et al. (2001) find a smoothing effect of the dynamics on the stress field in the precursory phase before a large event.

Huang et al. (2000) and Sammis and Smith (1999) have studied a simple cellular automaton model of earthquakes on a pre-existing hierarchical fault structure. The system is found to self-organize at large times in a stationary state with a power law Gutenberg-Richter distribution of earthquake sizes. The largest fault carries irregular great earthquakes preceded by precursors developing over long time scales and followed by aftershocks obeying an Omori's law. The cumulative energy released by precursors follows a time-to-failure power law with log-periodic structures, qualifying a large event as an effective dynamical (depinning) critical point. Down the hierarchy, smaller earthquakes exhibit the same phenomenology, albeit with increasing irregularities.

Mora et al (2000) and Mora and Place (2001) have used the lattice solid model (Place and Mora, 1999) to describe discontinuous elasto-dynamic systems subjected to shear and compression. They find accelerating seismic energy release in the lead-up to large earthquake events. They also document strong evidence that the stress-stress correlation length grows stronger as a large model earthquake is approached, in agreement with the prediction of the critical point concept (Mora and Place, 2001).

In a model of heterogeneous faults interacting through elastic coupling, Heimpel (1997) has found the existence of characteristic (but irregular) quake cycles that are broken up into four stages in agreement with the critical point hypothesis: (1) relaxation with aftershocks following the main-shock sequence of the previous cycle; (2) Self-organization in which stress and strength increase; (3) criticality characterized by large stress/strength fluctuations; (4) main shock which is accompanied by a rapid stress and strength drop.

Using a regional lithospheric model consisting of a seismogenic upper crust governed by the damage rheology of Lyakovsky et al. (1997) over a viscoelastic substrate, Ben-Zion and Lyakhovsky (2001) demonstrate the existence of accelerated seismic release prior to large model earthquakes, in the case when the seismicity preceding a given large earthquake has a broad frequency-size statistics. The accelerated seismicity is found to be accommodated both by increasing rates of moderate events and increasing average event size.

King and Bowman (2001) develop a model that is based solely on the decay of the stress shadow from a previous large event. In their model, the stress shadow, calculated using an elastic dislocation model of a great earthquake is relaxed linearly in time. At each time step, fractal noise is added to the stress and events are calculated for those areas above a failure threshold. Increases in event size and accelerated seismic release are produced by an increase in the number and size of patches above the failure threshold as the shadow decays.

Although each of these models appears to simulate observed seismic acceleration before great events, their formulations are quite different and their relation to the critical point hypothesis is not clear. In this paper, we attempt to clarify the critical point concept by pointing out that a critical point occurring as a function of time can be better seen as a so-called "finite-time singularity", a mathematical concept encompassing and generalizing that of a "critical point". Another advantage of this terminology is that the word "critical point" is loaded with many different meanings and has been used in distinct ways in several contexts and scientific communities. We exemplify the "finite-time singularity'' concept by simple mechanical models and then explore how models proposed to simulate regional seismicity give rise to a finite-time singularity.

**Finite-time singularities**

The mathematics of singularities is applied routinely in the physics of phase transitions to describe for instance the transformations from ice to water or from a magnet to a demagnetized state when raising the temperature, as well as in many other condensed matter systems. Such singularities characterize so-called critical phenomena and "critical points". This was the concept proposed initially by Sornette and Sammis (1995) for the description of large earthquakes. In these problems, physical observables such as susceptibilities, specific heat, etc., exhibit a singularity as the control parameter (temperature, magnetic field) approaches a critical value.

It may be useful to be more precise and view large earthquake as belonging to a larger class of singularities occurring in dynamical systems, which are spontaneously reached in finite time. Since earthquakes are organized in time with a great event terminating a "cycle", this seems to be a useful view point. Spontaneous singularities are quite common and have been found in many well-established models of natural systems, either at special points in space such as in the Euler equations of inviscid fluids (Pumir and Siggia, 1992; Bhattacharjee et al., 1995), in the surface curvature on the free surface of a conducting fluid in an electric field (Zubarev, 1998), in vortex collapse of systems of point vortices, in the equations of General Relativity coupled to a mass field leading to the formation of black holes (Choptuik, 1993; 1999), in models of micro-organisms aggregating to form fruiting bodies (Rascle and Ziti, 1995), or in the more prosaic rotating coin (Euler's disk) (Moffatt, 2000).

The simplest representative dynamical evolution equation leading to a finite-time singularity is

$$\frac{dE}{dt} = E^m \qquad \text{with } m>1. \qquad (2)$$

Its solution is

$$E(t) = E(0)\left(\frac{t_c - t}{t_c}\right)^{-\frac{1}{m-1}} \qquad (3)$$

where the critical time $t_c = (m-1)/[E(0)]^{m-1}$ is determined by the initial condition $E(0)$. The singularity results from the fact that the instantaneous growth rate $d\ln E/dt = E^{m-1}$ is increasing with E and thus with time. This can be visualized by studying the "instantaneous" doubling time, defined at the time interval $\Delta t$ necessary for $E(t)$ to double, i.e., $E(t + \Delta t) = 2E(t)$, if the growth rate was fixed constant at its value at time $t$. When the growth rate of $E$ increases as a power law of $E$, the doubling time decreases fast and the sequence of doubling time intervals shrinks to zero sufficiently fast so that its sum is a convergent geometrical series. The variable thus undergoes an infinite number of doubling operations in a finite time, which is the essence of a finite-time singularity. The growth of the growth rate captures a positive feedback of the variable $E$ on its growth rate. This positive feedback is the main ingredient for a "finite-time singularity". In the context of earthquake nucleation, Dieterich (1992) (see Sornette (2000) for a simple derivation) has shown that the usual slip- and velocity-dependent solid friction law introduced in the equation of elasticity gives rises to a finite-time singularity of the slip velocity, representing the transition from accelerated nucleating slip and the fast elasto-dynamical rupture regime. In this case, the positive feedback results from the fact that as the slip distance and slip velocity increase, the solid friction coefficient decreases, thus enhancing further slip.

We now discuss several other examples in which positive feedback leads to a finite-time singularity. We begin with models of crack growth and then move to models with distributed failure that are more suitable for regional seismicity.

**Positive feedback in crack growth**

A well-known example of a positive feedback occurs in the sub-critical crack growth usually represented by a power law equation

$$dL/dt = CK^p \qquad (4)$$

where $L$ is the crack length, $K$ is the stress intensity factor and $C$ and $p$ are material parameters. In slow early deformation phases, the stress corrosion exponent is typically in the range 2-5.

Using the standard result from continuum elasticity $K \propto L^{1/2}$, the sub-critical crack growth equation becomes

$$dL/dt = C'L^{p/2}. \tag{5}$$

Since $p/2>1$, this is nothing but eq.(2) with $m=p/2$, which gives the finite-time singularity for the crack length according to eq.(3). This singular behavior has been noted by several authors starting with Das and Scholz (1981) and has been used for instance to explain the power law distribution of fault length (Sornette and Davy, 1991). Ben-Zion and Lyakhovsky (2001) then convert it into an equation of the form (1) by using the relationship between seismic moment and fault length $M_0 \propto L^{2-3}$.

The finite-time singularity of the crack growth has recently been refined (Gluzman and Sornette, 2001) within a self-consistent theory of crack growth controlled by a cumulative damage variable d(t), dependent on stress history in an elastic medium, in the quasi-static regime where the sound wave velocity is taken as infinite. Depending upon the damage exponent $m$, which controls the rate of damage $dd/dt \propto \sigma^m$ as a function of local stress $\sigma$, two regimes are found. For $0<m<2$, the model predicts a finite time singularity. For $m \geq 2$, the rupture dynamics requires a regularization scheme in the form of a saturation of damage, a minimum distance of approach to the crack tip or a fixed stress maximum. This shows that the theory leading to a strong finite-time singularity has no continuous limit and is controlled by microscopic processes close to the crack tip. We stress that the mathematical existence of a finite-time singularity is nothing but the signature of a transition to another regime, here the transition to an elasto-dynamic rupture controlled by the finiteness of the speed of elastic shear waves.

Gluzman et al. (2001) have used the insight on the existence of finite-time singularities and their associated power law behavior to develop new theoretical formulas for the prediction of the rupture of systems which are known to exhibit a critical behavior, based solely on the knowledge of the early time evolution of an observable, such as the precursory acoustic emission rate or seismic rate as a function of time or of stress. From the parameterization of such early time evolution in terms of a low-order polynomial, the functional renormalization approach transforms this polynomial into a function that is asymptotically a power law. The value of the critical time $t_c$, conditioned on the assumption that it exists, can be determined from the knowledge of the coefficients of the polynomials. This prediction scheme shows a much stronger robustness and reliability with respect to the order of the polynomials and as a function of noise compared to direct power law fits.

**Positive geometrical feedback in laboratory creep rupture**

Consider the problem of so-called creep rupture (Krajcinovic, 1996) in which a rod is subjected to uniaxial tension by a constant applied axial force P. The cross section $A(t)$ of the rod is assumed to be a function of time. The problem is simplified by assuming that $A(t)$ is independent of the axial coordinate, which eliminates necking as a possible mode of failure. The creep deformation is assumed to be isochoric, i.e., the rod volume remains constant during the process. This provides a geometric relation between the cross-sectional area and length $A_0 L_0 = A(t)L(t) = \text{constant}$, which holds for all times.

The rate of creep strain $\varepsilon_c$ can be defined as a function of geometry as

$$\frac{d\varepsilon_c}{dt} = \frac{1}{L}\frac{dL}{dt} = \frac{1}{A}\frac{dA}{dt} \qquad (6)$$

showing that $\varepsilon_c = \ln(L(t)/L_0) = -\ln(A(t)/A_0)$, where $L_0 = L(t_0)$ and $A_0 = A(t_0)$ correspond to the undeformed state $\varepsilon_c = 0$ at time $t_0$. Assuming power-law creep

$$d\varepsilon_c/dt = C\sigma^\mu \qquad \text{with } \mu > 0, \qquad (7)$$

where $\sigma = P/A$ is the ratio of the applied force over the cross section of the rod. Eliminating $d\varepsilon_c/dt$ between (6) and (7) and using $\sigma = P/A$ leads to $A^{\mu-1}(dA/dt) = -CP^\mu$ whose solution is $A(t) = A(0)[(t_c - t)/t_c]^{1/\mu}$, where the critical failure time is given by $t_c = [A(0)/P]^\mu/(\mu C)$. The rod cross section thus vanishes in a finite time $t_c - t_0$ and as a consequence the stress diverges as the time $t$ goes to the critical time $t_c$ as $\sigma = P/A = [P/A(0)][(t_c - t)/t_c]^{-1/\mu}$. Physically, the constant force is applied to a thinner cross section, thus enhancing the stress, which in turn accelerate the creep strain rate, which translates into an acceleration of the decrease of the rod cross section and so on. In other words, the finite-time singularity results from the positive feedback of the increasing stress on the thinner geometrical cross section and vice-versa. This finite-time singularity for the stress can be reformulated as a self-contained equation expressed only in terms of the stress: $d\sigma/dt = C\sigma^{\mu+1}$, which is indeed of the form (1) since $\mu>0$.

**Positive feedback in a simple damage model**

Sammis et al. (1996) demonstrated that equation (1) follows from the most elementary form of damage mechanics. Their argument is closely analogous to that in the previous section and goes as follows. Let $V_i(t)$ be the volume fraction of intact rock at time $t$ and $V_d(t)$ be the volume fraction of damage rock such that $V_i(t) + V_d(t) = 1$. If the damaged rock is not able to support load, then the average local stress, $\sigma_{local}$, will be related to the regional remote stress, $\sigma_{remote}$, as $\sigma_{local} = \frac{\sigma_{remote}}{A_i} = \frac{\sigma_{remote}}{V_i}$ (since the average cross-sectional area fraction equals the volume fraction). If the average rate of damage production is proportional to the average local stress raised to some power $\beta>0$ (Norton's law), we have

$$\frac{1}{V_d}\frac{dV_d}{dt} = -\left(\frac{1}{1-V_i}\right)\frac{dV_i}{dt} = c\sigma_{local}^\beta = c\left(\frac{\sigma_{remote}}{V_i}\right)^\beta \qquad (8)$$

Equation (4) can be integrated to yield $t_c - t = cV_i^{\beta+1}$ (when $t \approx t_c$) where $V_i(t_c) = 0$. If the cumulative energy released at time $t$, $E(t)$, is proportional to the damaged volume $V_d(t)$ we have

$$E(t) = c_1 V_d(t) = c_1[1 - V_i(t)] = c_1\left[1 - c_2(t_c - t)^{1/(\beta+1)}\right] \qquad (9)$$

which has the form of equation (1). Note that the structure of (9) remains robust close to $t_c$ if $E(t)$ is an arbitrary monotonously increasing function of $V_d(t)$. The "finite-time singularity" at $t = t_c$ in equation (9) is the result of positive feedback in equation (8): since $\sigma_{local} = \frac{\sigma_{remote}}{V_i}$, we get $d\sigma_{local}/dt \propto \sigma_{local}^{\beta+2}$ which is indeed of the form (2) with β>0. The positive feedback comes from the geometrical feedback of the damage onto the local stress.

**Positive feedback in a damage model for regional seismicity**

Ben-Zion and Lyakhovsky (2001) calculate regional seismicity using the damage rheology developed by Lyakovsky et al. (1997). In their model damage creation is linked to the strain. In 1-D it takes the form $d\alpha/dt = c\varepsilon^2$ where $\alpha$ is the damage and $\varepsilon$ is the 1-D strain. Feedback in this case comes through the elastic constant $E = E_0(1-\alpha)$ which is an explicit function of the damage. Hooke's law is thus $\sigma = E_0(1-\alpha)\varepsilon$ which, at constant stress, leads to

$$\varepsilon = \frac{\sigma}{E_0}\left(\frac{t_c - t}{t_c}\right)^{-1/3} \quad \text{where} \quad t_c = \frac{E_0^2}{3C\sigma^2} \tag{10}$$

In (10), α=1 when t=$t_c$ which defines the critical time. Although the strain has a singularity at $t = t_c$, the cumulative Benioff strain does not. Following Ben-Zion and Lyakhovsky (2001), the elastic energy release rate can be written

$$\frac{dU}{dt} = \frac{d}{dt}\left(\tfrac{1}{2}\sigma\varepsilon\right) = \frac{\sigma^2}{2E_0}\left(\frac{t_c - t}{t_c}\right)^{-4/3} \tag{11}$$

which, when integrated to give the Benioff strain, gives

$$\int\left(\frac{dU}{dt}\right)^{1/2} dt \propto (t_c - t)^{1/3} \tag{12}$$

consistent with the observed values. This model is another example of a finite-time singularity caused by positive feedback. Differentiating Hooke's law and using $d\alpha/dt = c\varepsilon^2$ gives $(d\varepsilon/dt) = (cE_0/\sigma)\varepsilon^4$ which has the form of (2) and leads to the finite-time singularity in $\varepsilon$.

**Positive feedback in a percolation model of regional seismicity**

In the previous examples, the finite-time singularity was the result of positive feedback in the failure process at constant stress. It is also possible to have a finite-time singularity in a heterogeneous system driven by increasing stress, as when the stress shadow from a prior great earthquake decays due to tectonic loading. Imagine that the heterogeneity of the crust is accounted for by stochastic rupture thresholds $\sigma_{th}(r)$, where $r$ is position, distributed according to some distribution with only short-range spatial correlation. Let us assume that the last greatest earthquake has cast a stress shadow $-\sigma_0$ over a very large area $S$ and let us neglect for the time being its spatial dependence and assume that it is uniform in space. Tectonic loading then progressively increases the stress everywhere uniformly at the rate $d\sigma_{tect}/dt$. The stress-to-rupture at an arbitrary point $r$ at time $t$ is thus

$$\Delta\sigma(r) = \sigma_{th}(r) - \sigma_0 + \left(\frac{d\sigma_{tect}}{dt}\right)t \tag{13}$$

We assume that only those points $r$ such that $\Delta\sigma(r) > 0$ are activated and can rupture as earthquakes. The other points with negative stress-to-rupture are inactive. The field $\Delta\sigma(r)$ can be represented as a highly corrugated random mountainous landscape in which the altitude $\Delta\sigma(r) = 0$ defines the boundary between seismic active regions ($\Delta\sigma(r) > 0$) and inactive domains ($\Delta\sigma(r) < 0$). The altitude $\Delta\sigma(r) = 0$ can be envisioned as a "sea level". Immediately after the large earthquake, most of the points $r$ have $\Delta\sigma(r) < 0$, which means that only a few islands emerge above "sea level". Ignoring aftershocks, the seismic activity is very weak, localized and incoherent. As time goes on and tectonic loading increases, the "mountainous landscape" $\Delta\sigma(r)$ is progressively lifted up. As a consequence, a larger and larger distribution of islands emerges above "sea level", until a point when macroscopic "continents" appear. This process is nothing but the percolation model governed by the control parameter defined as the fraction $p$ of territory above "sea level". At early time after the great event, $p$ is very small and the stress landscape is below percolation. As time increases, $p$ increases approximately linearly with time, until it reached a critical value $p_c$, called the percolation threshold, corresponding to the appearance of very large clusters of connected points above "sea level".

Percolation theory (Stauffer and Aharony, 1994) provides precise prediction on the distribution of cluster sizes $n(s)$, i.e., the distribution of earthquake sizes in this simple model. It is known that, sufficiently close to $p_c$, the number of clusters of size $s$ per unit area is given by

$$n(s) \propto s^{-\tau} \exp[-s/s^*(p)] \qquad (14)$$

where the typical maximum cluster size diverges as

$$s^*(p) \propto |p - p_c|^{1/\sigma} \qquad (15)$$

as $p$ goes to the critical point $p_c$. $s^*(p)$ is nothing but the correlation length of the system. $\tau$ and $\sigma$ are two critical exponents which depend upon the dimension of the system. In two dimension, $\tau=187/91$ and $\sigma=36/91$ while in high dimensions or in mean-field theory $\tau=5/2$ and $\sigma=1/2$. Note that, at the critical point $p = p_c$, the distribution of earthquake size becomes exactly self-similar according to a pure power law. By definition, $\sum_s sn(s) = p$. The average size of earthquakes (clusters) is then given by

$$\langle s \rangle = \frac{\sum_s s^2 n(s)}{\sum_s sn(s)} \propto |p - p_c|^{-\gamma} \qquad (16)$$

where $\gamma$ is another critical exponent, related to $\tau$ and $\sigma$ by the scaling law: $\gamma = (3-\tau)/\sigma$. The released seismic energy is usually taken as the 3/2 power of the rupture area $s$. This shows that the average seismic energy released per earthquake is expected to increase as the power law $|p - p_c|^{-3\gamma/2}$ as the critical point is approached. This simple stress percolation model thus predicts an accelerated increase with time of the typical largest earthquake $s^*(p)$ and of the average seismic energy $\langle s \rangle^{3/2}$ according to the law of finite-time singularity. Equation (16) can indeed be re-written as is $\frac{d\langle s \rangle}{dt} = \langle s \rangle^\lambda$ with $\lambda=(\gamma+1)/\gamma>1$. This expression can also be recovered from a mapping of the random point percolation problem on a version of the forest fire model using a cluster analysis (Turcotte, 1999).

There is, however, a problem. For random percolation, the power γ is always greater than one (e.g., in 2D, $\gamma = [(3-\tau)/\sigma] \approx 3$) so that not only is average seismic energy in a given time interval (i.e., the energy rate) singular, but so is its integral, the cumulative energy (or cumulative Benioff strain). Such is not the case when the percolating topography is correlated. In particular, it is well-known that rupture can be modeled as correlated percolation (Sahimi, 1998; Johansen and Sornette, 2000): the correlation strength continuously evolves in a system progressively brought to failure via the action of the stress field that provides an additional channel of communication on top of the sole connectivity defined in standard percolation. A conceptual model of such correlated percolation mediated by the existence of an additional field such as the stress is to consider spins interacting with ferromagnetic interactions (playing the role of the stress field) on a percolating lattice. In this case, the exponents can be calculated and one finds that the exponent $\sigma$ is changed into 2 rather than of the order of 1/2 or 1/3 and the exponent $\gamma$ becomes less than 1 (Stauffer and Aharony, 1994; p. 146). In this class of correlated percolation models with an addition field, the probability to be above sea-level is a function of past events due to stress redistribution. In this sense, stress redistribution gives positive feedback in the percolation model where p is not only growing globally as a function of time but is modified locally due to stress redistribution by prior events. For the cumulative Benioff strain which is proportional to the integral over time of the average cluster size, this correlation in time with path-dependent history leads to γ<1 and therefore to a finite-time singularity of the form (1).

**Positive feedback in a stress-shadow model for regional seismicity**

In the previous model, the accelerating seismicity and finite-time singularity emerge as the direct consequence of two ingredients: (1) the power law dependence of the correlation length as a function of distance to the critical point and (2) the sweeping of the control parameter *p* which increases monotonically with time (see Sornette, (1994) for a general discussion of sweeping of control parameters in critical point models). In this model, the accelerated seismicity is expressing the growth of correlation of stress within the system quantified by the size of the clusters above the rupture threshold. King and Bowman (2001) modify the uniform stress shadow by calculating the spatial redistribution of stress predicted by elastic dislocation theory, to which they added fractal noise. Their tectonic loading is not uniform but also consists in decreasing progressively the strength of the elastic dislocation equivalent to the great earthquake. In addition, they allow the noise to be redistributed at each time step. When they increased this regional stress field linearly in time to simulate tectonic loading, and applied a threshold failure criterion, their model seismicity was well fit by equation (3). In their model, $t_c$ is the time at which tectonic loading has eliminated the stress shadow. We now give an analytic solution to this problem with the goal of deriving equation (1).

Consider the section shown in Figure 1 orthogonal to the fault on which the great event last occurred. This event is supposed to cast a stress shadow, which, for simplicity, we take as a negative scalar field $\sigma_{sh}(r)$ that is only function of the distance r from the fault:

$$\sigma_{sh}(r) = -\sigma_0 \frac{h^d}{\left(r^d + h^d\right)} \qquad (17)$$

where $-\sigma_0$ is the stress drop on the fault, *h* is the depth of the fault, and *d* is the dimension of faulting (*d*=2 for fracture through a plate, *d*=3 for a half-space). To this stress shadow we add random stress fluctuations to represent strength heterogeneity and stress perturbations associated with prior events. We will assume a constant failure threshold. This is equivalent to assuming random failure thresholds and spatially smooth stress level as in the previous section. The second ingredient of the model is a simulation of tectonic loading by decreasing the strength of the dislocation linearly with time according to $\sigma_0(t) = \sigma_0(0) - d\sigma_{tect}/dt$. The total average stress applied at a point $r$ from the fault is thus

$$\sigma(r,t) = -\left(\sigma_0(0) - \frac{d\sigma_{tec}}{dt}t\right)\left[\frac{h^d}{r^d + h^d}\right] \qquad (18)$$

where the origin of time is taken at the time of the last great earthquake. As the third ingredient of the model, we assume that intermediate sized earthquakes only become possible when the average stress level reaches some threshold $-\sigma^* < 0$ corresponding to the percolation threshold discussed above. At time $t$, we can thus define a boundary $r^*(t)$ such that all points $r > r^*(t)$ are above the percolation threshold and can generate intermediate events while all points $r < r^*(t)$ are still in the shadow below the percolation threshold and can only produce small events (see Fig.1). The transition distance $r^*(t)$ is such that

$$\sigma(r^*,t) = -\sigma^* = -\left(\sigma_0(0) - \frac{d\sigma_{tec}}{dt}t\right)\left[\frac{h^d}{\left(r^{*d} + h^d\right)}\right] \qquad (19)$$

which gives

$$r^*(t) = h\left[\left(\frac{t_c - t}{t_c - t^*}\right) - 1\right]^{1/d} = h\left[\left(\frac{t^* - t}{t_c - t^*}\right)\right]^{1/d} \qquad (20)$$

where $t_c \equiv \sigma_0(0)/(d\sigma_{tec}/dt)$, the time at which the tectonic stress has reloaded the main fault so that it can rupture again. Time $t^*$ is defined at the time at which $\sigma(0,t^*) = -\sigma^*$ and the entire shadow has reached the percolation threshold. Hence $r^*(t^*) = 0$ (see (20)). We also may write $t_c - t^* = \sigma^*/(d\sigma_{tec}/dt)$. Note that $r^*(t)$ goes to zero in finite time as $t \to t^*$ with a singular behavior having exponent $1/d$). Thus, new area reaches the percolation threshold at a rate proportional to $-(dr^*/dt)$.

If the total area were constant, the cumulative number of earthquakes would simply grow linearly with time. In the present case, the active area grows with time. We assume that when an elementary domain sees its total applied stress $\sigma(r)$ given by (16) reach the percolation threshold $\sigma^*$, a sudden burst of earthquake energy is released as a result of the rapidly growing correlation. After this brief burst, the seismic activity assumes a constant background level. As a consequence, the cumulative energy released by earthquakes, $E(t)$ obeys the following differential equation

$$\frac{dE}{dt} = aL(R - r^*(t)) - gL\frac{dr^*}{dt} \qquad (21)$$

where $a$ is the constant background seismicity of the region freed from the large earthquake shadow per unit area, $L$ is the length of the fault and $g$ quantifies the intensity of the burst of seismic activity triggered by the percolation. $R$ is introduced as a convenient regularization, which has the physical meaning of the total size of the region influenced by the great earthquake. The results are independent on its specific value. We stress that the second term -gL dr*/dt of (21) is present as soon as the seismicity at the border of the shadow region is distinct from that outside the shadow. This can for instance take into account overshoot of the earthquakes suddenly freed and/or hysteresis of the mechanical deformations.

Neglecting the second term in the r.h.s. and for constant $r^*$, equation (21) becomes $dE/dt = aL(R - r^*)$ whose solution is $E(t) = aL(R - r^*)t$, corresponding to the usual linear-in-time cumulative number of earthquakes.

Close to $t^*$, the integration of the first term, $aL(R - r^*(t))$, gives a non-singular term whose leading behavior is proportional to time t. Close to $t^*$, it can be approximated by a constant $E_0$ since it varies much more slowly and smoothly than the other term. More interestingly, the second term $-gL(dr^*/dt)$ gives rise to the singular behavior (20) with infinite accelerating slope close to $t_c$. Putting everything together, we get finally that the cumulative energy release is governed by

$$E(t) = E_0 - ghL\left(\frac{t^* - t}{t_c - t^*}\right)^{1/d} \quad \text{for } t \to t^*. \tag{22}$$

It is interesting that stress on the main fault does not fully recover at $t=t^*$ but requires an additional time increment $t_c - t^* = \sigma^*/(d\sigma_{tec}/dt)$ that depends on the percolation threshold $\sigma^*$. Expression (19) has exactly the form (1) with $m=1/d$ and $B = -ghL/(t_c - t^*)$. This simple calculation provides maybe the simplest example of how an accelerating power law behavior culminating at a finite-time singularity may emerge from a simple geometrical mechanism. The exponent $m=1/d$ of the singular behavior is solely controlled by the $1/r^d$ dependence at large $r$ of the stress shadow $\sigma_{sh}$ given by (14). If $d=2$ for a large rupture that spans the brittle crust then $m=1/2$, if $d=3$ for a smaller rupture contained in the crust, then $m=1/3$.

The functional form of (22) is unchanged when including explicitly heterogeneous residual stress field and stress thresholds which vary from point to point in a random way. A statistical theory extending the previous calculation can be formulated which amounts to summing over the probability distribution of threshold and over space. The exponent $1/d$ is modified into $b+1/d$ if the distribution of stress thresholds is of the form $P(\sigma_{th}) \propto \sigma_{th}^b$, where $0 \leq b$ for small $\sigma_{th}$, i.e., if there are few elements with weak rupture thresholds close to 0. The larger $b$ is, the stronger is the rock system as there are fewer and fewer weak elements. Therefore, in this theory, the finite-time singularity relies on the existence of a significant fraction of weak faults or cracks (such that b is close to 0) which are destabilized as the stress shadow disappears.

**Discussion**

Models that have been proposed to describe accelerating seismicity before large earthquakes fall into two general classes: those based on mechanical feedback and those based on the decay of the stress shadow from a prior event. In the former, the power law acceleration of regional seismicity is due to positive feedback in the failure process at constant large-scale stress. This feedback can be the result of stress transfer from damaged to intact regions, or it can result from the effect of damage in lowering the local elastic stiffness. This result into finite-time singularities.

Models based on the recovery of a stress shadow due to tectonic loading and stress transfer are quite different. Accelerating seismicity results from the increasing stress as the shadow is eroded by tectonic loading. In this case, a finite-time singularity can result from either correlated percolation (describing stress transfer between intermediate size events) in a uniformly rising stress field, or from the geometrical shrinking of the $r^{-d}$ stress perturbation coupled with random percolation. In the latter case, the finite-time singularity results in a "front" of intermediate sized events that propagate toward the main fault as area closer to the fault progressively reach the percolation threshold.

Although both classes of models predict accelerating seismicity and growth of event size before a great earthquake, they predict different spatial patterns for this precursory seismicity that may offer an observational way to distinguish between them. Models based on feedback in the failure process at constant stress predict rapid shear localization with the largest precursory event occurring in or near the fault plane of the of the ultimate event. Models based on the decay of stress shadows, on the other hand, predict that the largest precursors should develop first at the fringes of the shadow and gradually move toward the plane of the great event. Current observations tend to find intermediate precursory events at large distances (Bowman et al., 1998; Jaume and Sykes, 1999), but a quantitative test of the two classes of models has yet to be performed.

Finally, we end with a somewhat speculative figure that tends to support a mechanism based on the decay of stress shadows. Amos Nur (private communication) has observed that strike slip earthquakes in the Mojave Desert have been occurring progressively closer to the San Andreas Fault since 1857. In Figure 2, the ordinate is the distance of each event from the closest point on the San Andreas Fault while the abscissa is the elapsed time since the last great event on that portion of the fault (1857 to the north, 1812 to the south). Coordinates of the San Andreas fault were taken from the USGS data base at (http//geohazards.cr.usgs.gov/eq/faults/fsrpage07.shtml). Earthquake epicenters and magnitudes were taken from the "Ellsworth Catalog" of historical seismicity that can be found at (http//pasadena.wr.usgs.gov/info/cahist_eqs.html) and from the map at the Southern California Earthquake Center (SCEC) data center (http//www.scecdc.scec.org/clickmap.html). Although there are not many events, there is a monotonic decrease in distance with time.

**Acknowledgements**: This work was partially supported by NSF-DMR99-71475 and the James S. Mc Donnell Foundation 21st century scientist award/studying complex system (DS).

Figure Captions

Figure 1. Stress shadow following a great earthquake. The average stress varies as $1/r^d$ as indicated by the dashed curve. Areas where the stress is greater than σ=0 can support earthquake rupture. When the average stress reaches -σ* the areas above σ=0 are at the percolation threshold. Large precursory event are much more likely for r ≈ r* where $\sigma(r^*,t) \equiv \sigma^*$.

Figure 2. Distance of intermediate sized (M>4.8) strike-slip events in the Mojave Desert from the San Andreas Fault as a function of time since the last great earthquake (1859 to the north, 1812 to the south). Aftershocks have been excluded.